\documentclass{article}                                                    

\usepackage{graphicx}

\newcommand{\eefourpi}{\ensuremath{e^+e^- \to
    \pi^+\pi^-\pi^+\pi^-}}
\newcommand{\fourpi}{\ensuremath{\pi^+\pi^-\pi^+\pi^-}}
\newcommand{\pne}{\ensuremath{\pi^{+}\pi^{-}\pi^{0}\pi^{0}}}

\date{}
\title{STATUS OF EXPERIMENTS AND RECENT RESULTS FROM CMD-2 DETECTOR AT
  VEPP-2M}
\author{A.I. Milstein\\
  (CMD-2 collaboration)\\
{\em Budker Institute of Nuclear Physics, Novosibirsk, 630090,
  Russia}}

\begin{document}

\baselineskip=11.6pt

\maketitle 

\begin{abstract}
  The Cryogenic Magnetic Detector (CMD-2) is  shortly described.
Preliminary results for the cross sections of $e^+$ $e^-$ annihilation
into hadrons and leptons are presented in the c.m. energy range from 0.37
to 1.39 GeV. The total integrated luminosity of about 26
$\mbox{pb}^{-1}$ has been
collected. The new results for the $\rho$  and $\omega$ meson parameters are
reported. The major decay modes of the $\phi$ meson as well as its rare decays
have been observed.
\end{abstract}

\baselineskip=14pt

\section{Introduction}
The investigation of the reaction of $e^+e^-$ annihilation into
hadrons at low energies accounts about thirty years history of the
experimental studies. Nevertheless, the understanding of the field is
still rather far from the completeness. More precise measurements of the
$\rho$-, $\omega$- and $\phi$-meson parameters are needed as well as
properties of the continuum which provide unique information about
interaction of light quarks and spectroscopy of their bound states.
The knowledge of the total cross section of $e^+e^-$ annihilation into
hadrons at low energies and
the magnitude of the exclusive cross sections
is also necessary for precise calculations of various
quantities. One of them is the strong interaction contribution
into anomalous magnetic momentum of muon $(g-2)_{\mu}$.
Tab.~\ref{tab-hcgm} shows contributions of various channels
into $(g-2)_{\mu}$ (details of calculations can be found in\cite{ej}
and\cite{bw}).
One can see, that the main contribution (about 87\%) comes from the
energy region below 1.4 GeV and the dominant contribution (71\%)
in this region is from the channel $e^+e^-\to\pi^+\pi^-$.

\begin{table}[h]
\caption{\it Hadronic contributions to the muon
  $(g-2)_{\mu}$}\label{tab-hcgm}
\begin{center}
\begin{tabular}{|c|c|c|c|c|}
  \hline
   & \multicolumn{2}{|c|}{$< 1.4$ GeV} &
  \multicolumn{2}{|c|}{$>1.4$ GeV} \\
  \cline{2-5}
  Mode & $\times 10^{-10}$ & \% & $\times 10^{-10}$ & \% \\
  \hline
  $\pi^+\pi^-$           & 502.2 & 71  & 0.78  & 0.1 \\
  $\pi^+\pi^-\pi^0$      & 50.96 & 7.2 & 0.67  & 0.1 \\
  $K^+K^-$               & 20.62 & 2.9 & 2.21  & 0.7 \\
  $K_S K_L$              & 0.76  & 0.1 & 14.22 & 2.0 \\
  $\pi^+\pi^-\pi^0\pi^0$ & 10.78 & 1.5 & 8.54  & 1.2 \\
  $\pi^+\pi^-\pi^+\pi^-$ & 5.16  & 0.7 & 10.22 & 1.5 \\
  $\pi^+\pi^-\pi^+\pi^-\pi^0\pi^0$ & & & 5.09  & 0.7 \\
  $\pi^+\pi^-\pi^+\pi^-\pi^0$ & 0.31 & & 2.53  & 0.4 \\
  $K_S K^{\pm}\pi^{\mp}$ &       &     & 0.951 & 0.1 \\
  $K^+K^-\pi^+\pi^-$     &       &     & 0.815 & 0.1 \\
  $K^{\ast}K^{\pm}\pi^{\mp}$ &   &     & 0.692 & 0.1 \\
  \hline
\end{tabular}
\end{center}
\end{table}

The energy behavior of cross section of the process $e^+e^-\to
hadrons$ at low energies is rather complicated. 
It is characteristic of various resonances ($\rho$, $\omega$, $\phi$
and their recurrences) and onsets of the separate hadronic
channels. Thus, to determine the value of the total cross section of
$e^+e^-$ annihilation into hadrons, one need to measure individual
channels one by one and study decay modes of $\omega$ and $\phi$
mesons.

These physical tasks became the goal of the general-purpose detector
CMD-2\cite{CMD} which has been running at the VEPP-2M
$e^+e^-$ collider\cite{vepp} in Novosibirsk since 1992 studying
the c.m. energy range from threshold of hadron production to 1.4 GeV.

   The CMD-2 detector is described in detail elsewhere\cite{CMD}.
It is a general purpose detector consisting of a drift chamber
(DC)\cite{DC} with
about 250 $\mu$ resolution in the plane transverse to the beam axis and
multiwire proportional chamber (ZC)\cite{ZC}
with an accurate measurement ($\sim$0.5
mm) of the z-coordinate of particle track along the beam direction.
Both chambers are inside a thin (0.38 $X_{0}$) superconducting solenoid
with a field of 1 T.

   The barrel calorimeter placed outside of the solenoid and
consists of 892 CsI crystals\cite{CsI} of  $6\times 6\times 15$ $cm^{3}$
size. The crystals are arranged in eight octants. The light readout is
performed by PMTs. The
energy resolution is about $8\%$ for photons with the energy more than
100 MeV. Both azimuthal and polar angle resolution is about 0.02 radian.

   The endcap calorimeter\cite{BGO} consists of 680 BGO
crystals of $2.5\times 2.5\times 15$ $cm^{3}$ size.
The light readout is performed by
vacuum phototriodes placed on the crystals. The energy and angular resolution
were found to be $\sigma_{E}/E = 4.6\%/\sqrt{E(GeV)}$ and $\sigma_{\phi,
\theta} = 2\cdot10^{-2}/\sqrt{E(GeV)}$ radians respectively. The solid
angle covered by both parts of the calorimeter is about 96\% of 4$\pi$.

The muon range system\cite{rs} consists of two double layers of the
streamer
tubes operating in a self-quenching mode and is aimed to separate pions and
muons. The inner and outer parts of this system are arranged in 8 modules each
and cover 55\% and 48\% of the solid angle respectively.

\section{Measurement of the pion form factor and $\rho$, $\omega$ meson 
parameters}

\begin{figure}[ht]
\includegraphics[width=0.47\textwidth]{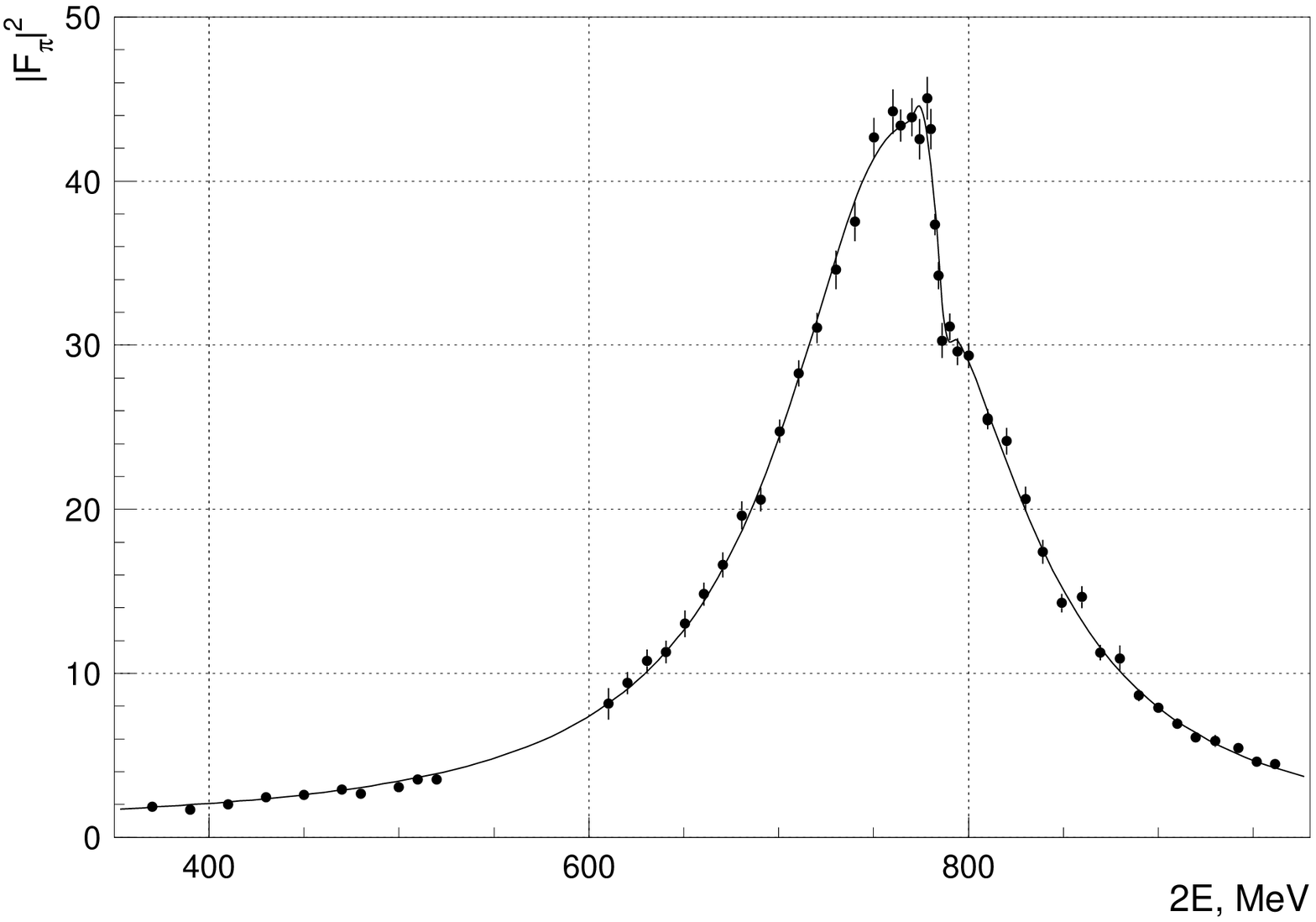}
\hfill
\includegraphics[width=0.47\linewidth]{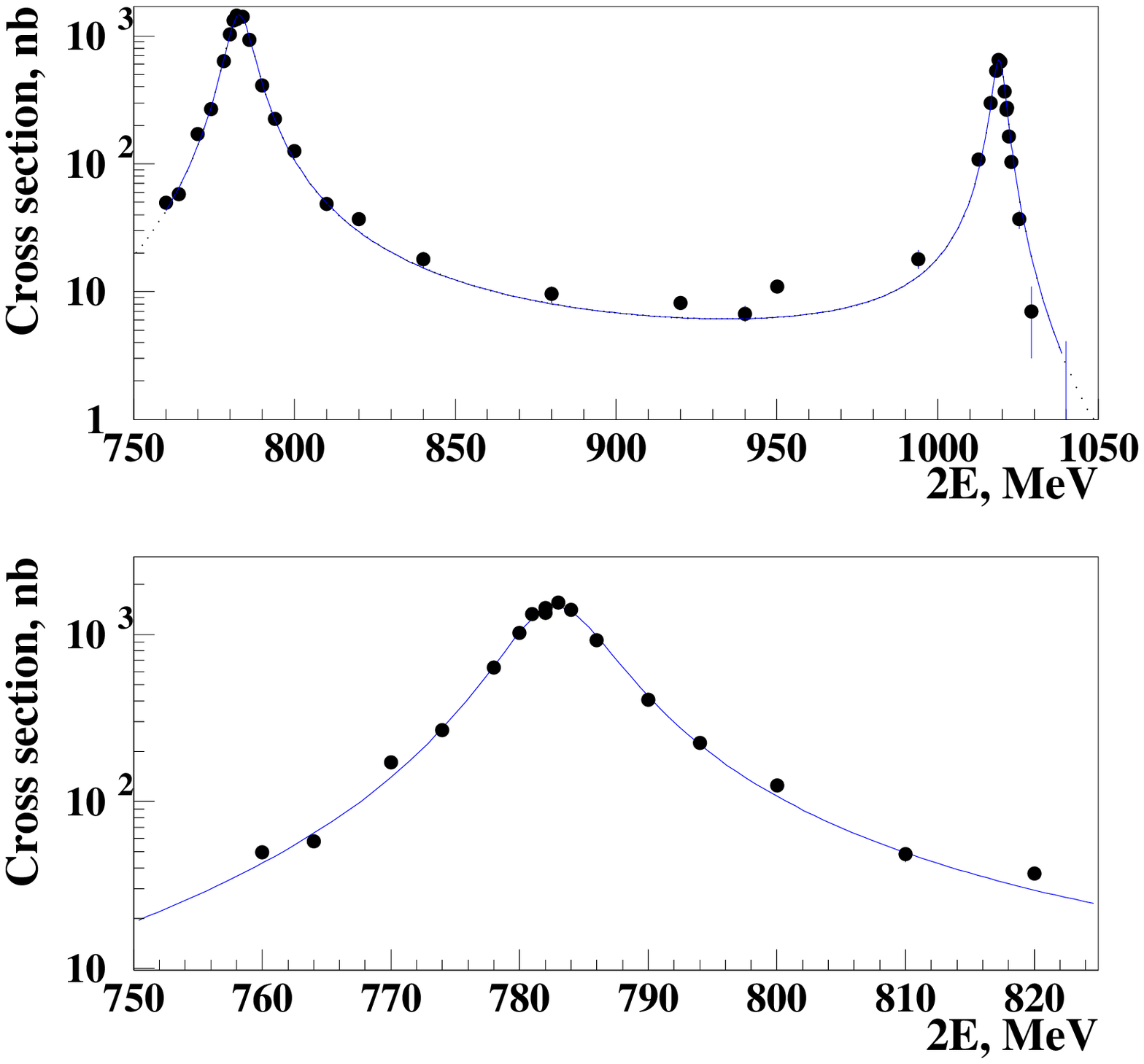}
\parbox[t]{0.47\textwidth}{\caption{\it The pion form factor vs
    c.m.energy and the fit with  Gounaris-Sakurai parametrisation
    .}\label{formfactor}}
\hfill
\parbox[t]{0.47\textwidth}{\caption {\it The upper figure - cross
    section     vs c.m.energy and common fit for the $\omega,\phi$
    mesons. The     lower figure - cross section in the energy range
    around $\omega$-meson.}\label{omega}}
\end{figure}

  A large data sample of about 2 
million $e^+e^-\to\pi^+\pi^-$ events was collected by CMD-2 
detector in the energy range from 0.360 to 1.370 GeV.
Analysis is completed for 10\% of the data only. 
The beam energy was measured by the resonance 
depolarization technique at almost all energies. The pion form factor
presented in fig.\ref{formfactor} is based on the data sample at 53 energy 
points in the 
energy range from 0.37 to 0.96 MeV. The obtained $\rho$ meson parameters based
on Gounaris-Sakurai parametrization were found to be:\\
{\boldmath 
$M_{\rho}$ = $775.28 \pm 0.61 \pm 0.20$ MeV, 
$\Gamma_{\rho}$ = $147.70 \pm 1.29 \pm 0.40$ MeV,\\
$\Gamma_{\rho\to e^+e^-}$ =$6.93\pm 0.11\pm 0.10$ keV, \\
$Br(\omega\to\pi^+\pi^-)$=$(1.31\pm0.23\pm0.02)\%$.}\\
Here and below the first errors are statistical and the second are 
systematic.
More details about the pion form factor can be found 
in\cite{form-factor}.  

   The energy range around the $\omega$ meson has been scanned at 13 energy 
points with a total integrated luminosity of about 1.5 $\mbox{pb}^{-1}$, but 
the detailed analysis was performed for $10\%$ of the data. 
The $\omega$ meson
parameters were measured with high accuracy using the $\omega \to 
\pi^+\pi^-\pi^{0}$ decay mode. 
The following parameters have been obtained from the fit:\\
{\boldmath 
$M_{\omega}$ = $782.71\pm 0.07\pm 0.04$ MeV,
$\sigma_{0}$ = $1482\pm 23\pm 25$ nb,}\\
{\boldmath 
$\Gamma_{\omega}$ = $8.68\pm0.23\pm 0.10$ MeV,
$\Gamma_{e^+e^-}$ = $0.605\pm 0.014\pm 0.010$ keV.}\\   
The common excitation curve for the $\omega$ and $\phi$ meson is presented in 
fig.\ref{omega}.

\section{Measurements of $\phi$ meson parameters}

\begin{figure}[ht]
\begin{center}
\includegraphics[width=0.47\textwidth]{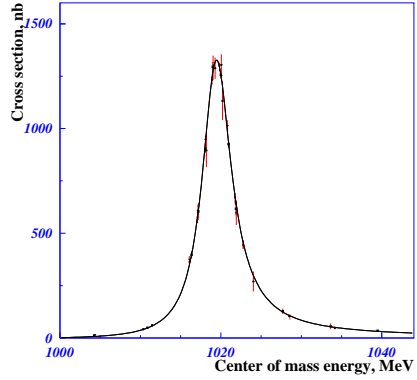}
\end{center}
\caption{\it Experimental cross section and $\phi$ meson excitation
  curve   in the channel $e^+e^-\to K^{0}_{L}
  K^{0}_{S}$}\label{phi-kskl}
\end{figure}

  The $\phi$-meson parameters were measured using data on the four major decay 
modes of $\phi\to$ $K_{S}K_{L}$, $K^{+}K^{-}$, $3\pi,$ $\eta\gamma$. 
The first 
results based on a relatively small integrated luminosity of about 300 
$\mbox{nb}^{-1}$ were published in\cite{phi}. The new more precise
results were  
obtained for the channel $\phi\to K^{0}_{L}K^{0}_{S}$ when 
$K^{0}_{S}$ decays into a $\pi^+\pi^-$\cite{phikskl}. The data sample
was collected in four scans of the energy range from 984 to 1040
MeV with the integrated luminosity of 2.37 $\mbox{pb}^{-1}$ and contains 
$2.97\times 10^5$ of selected $K^{0}_{L}
K^{0}_{S}$. Fig.\ref{phi-kskl} shows energy dependence of the cross 
section of the reaction $e^+e^-\to K^{0}_{L} K^{0}_{S}$ and
the excitation curve of the $\phi$ meson. The following parameters
have been obtained from the fit:
{\boldmath
$\sigma_{0}(\phi\to K^{0}_{L} K^{0}_{S})=1312 \pm 7\pm 33$ nb,\\
$M_{\phi}=1019.470\pm 0.013\pm 0.018$ MeV, $\Gamma_{\phi}=4.51
\pm 0.04\pm 0.02$ MeV, \\
$\Gamma_{\phi\to ee}\cdot Br(\phi\to K^{0}_{L}K^{0}_{S})$ = 
$(4.181\pm 0.024\pm 0.084)\times 10^{-4}$ MeV}.

\section{Study of $\phi\to\eta\gamma\to\pi^+\pi^-\pi^0\gamma$
decay}

\begin{figure}[ht]
\includegraphics[width=.47\textwidth]{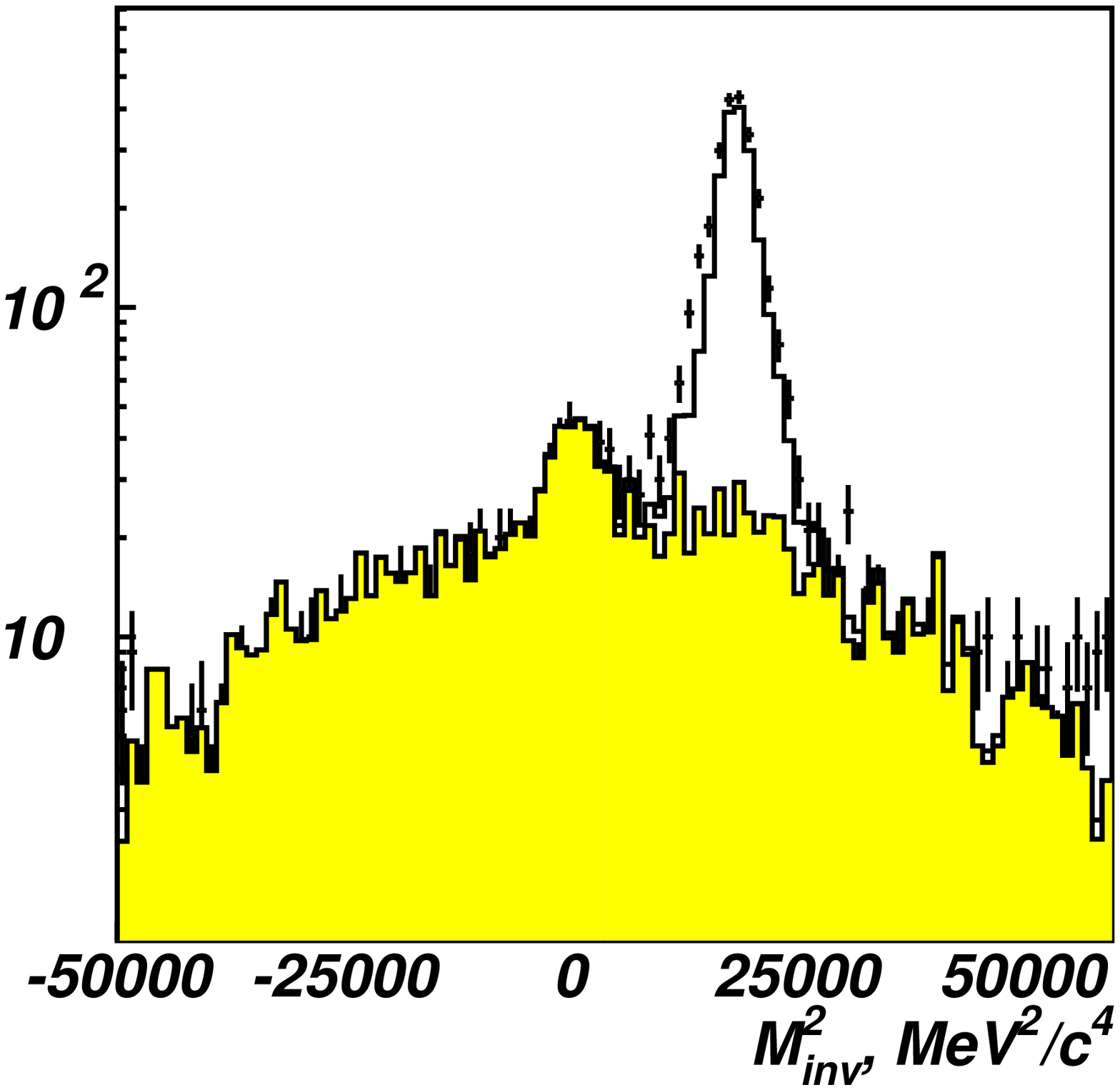}
\hfill
\includegraphics[width=.47\textwidth]{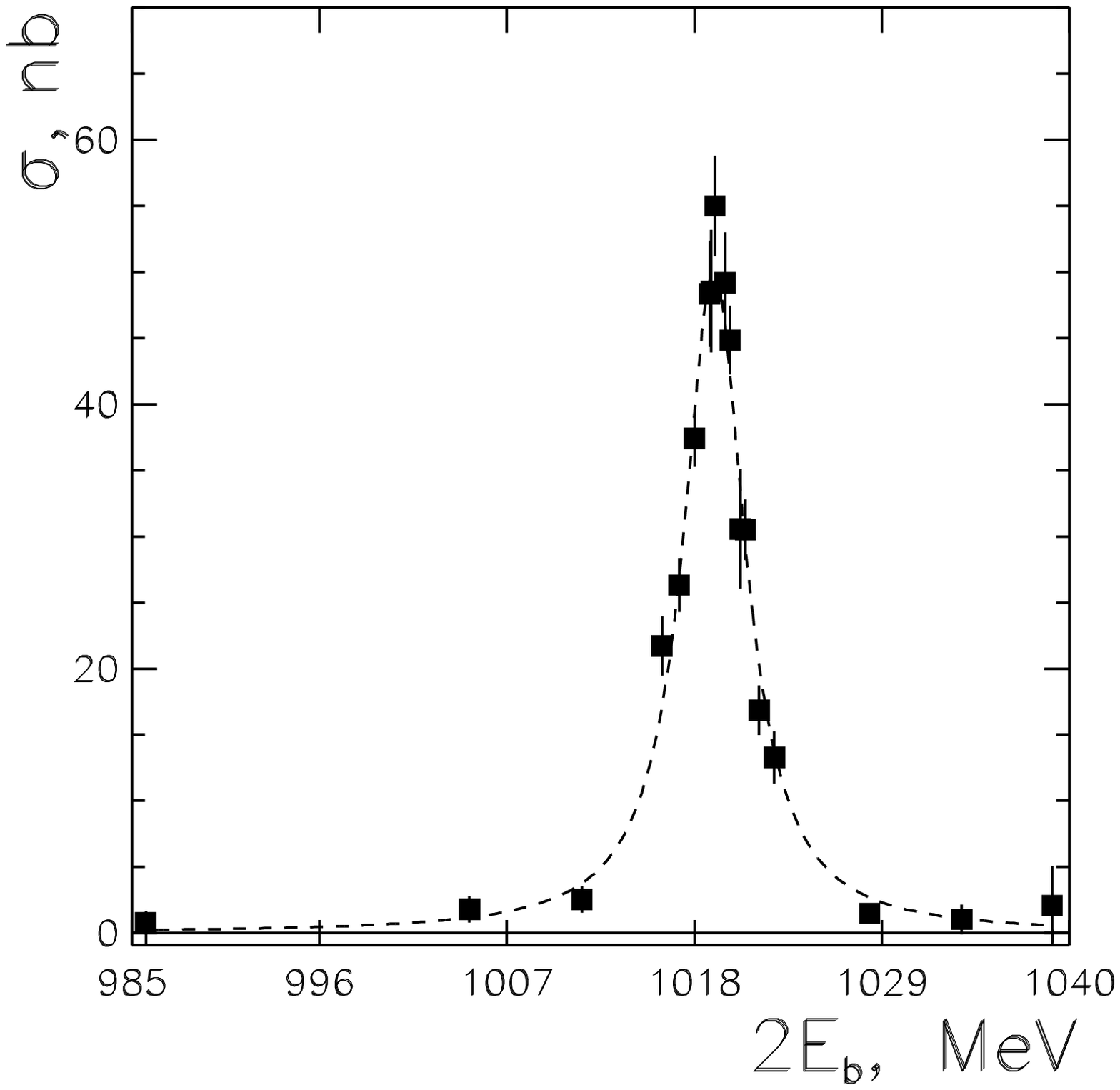}
\\
\parbox[t]{.47\textwidth}{\caption{\it {\bf $M^2_{inv}$}
    distribution. Points with errors present the data. The hatched
    histogram is the sum of such distributions for simulation of
    background processes, solid line histogram is the simulation of
    $\eta \gamma \to \pi^+ \pi^- \pi^0 \gamma$ together with
    background simulation} \label{fig:exp}}
\hfill
\parbox[t]{.47\textwidth}{\caption{\it The cross-section $\sigma(e^+e^-
    \to \phi \to \eta\gamma)$ with fit function}
  \label{fig:sigphi}}
\end{figure}

   The radiative magnetic dipole transition of $\phi$ into $\eta$ has been 
studied with the integrated luminosity of about 1.9 $\mbox{pb}^{-1}$. Events
with two charged particles and one recoil photon with the energy more 
than 250 MeV were selected. The direction of the recoil photon should be close
to the opposite direction of two charged pions. 
The reconstructed invariant mass of all other photons in this system
(fig.\ref{fig:exp})
forms a peak near the $\pi^{0}$ mass or 
near zero corresponding to the events of the $\eta$ decay into 
$\pi^+\pi^-\gamma$. The small fraction of the background comes from 
$\phi$ decays into $\pi^+\pi^-\pi^{0}, \omega\pi^{0},$ 
$K^{0}_{L}K^{0}_{S}$ and was subtracted
according to the simulation results.
Fig.\ref{fig:sigphi} shows the energy behavior of the cross section
of the process $e^+e^- \to \phi \to \eta\gamma$.
Using the branching ratio for the $\phi\to e^+e^-$ decay from
PDG\cite{PDG}, the following value for branching ratio has been
determined:
{\boldmath
$Br(\phi\to \eta\gamma) = (1.18\pm 0.03\pm 0.06)\times 10^{-4}$.}

\section{Process $e^+e^-\to \eta\gamma \to
  \pi^0\pi^0\pi^0\gamma$}

\begin{figure}[ht]
\includegraphics[width=0.47\textwidth]{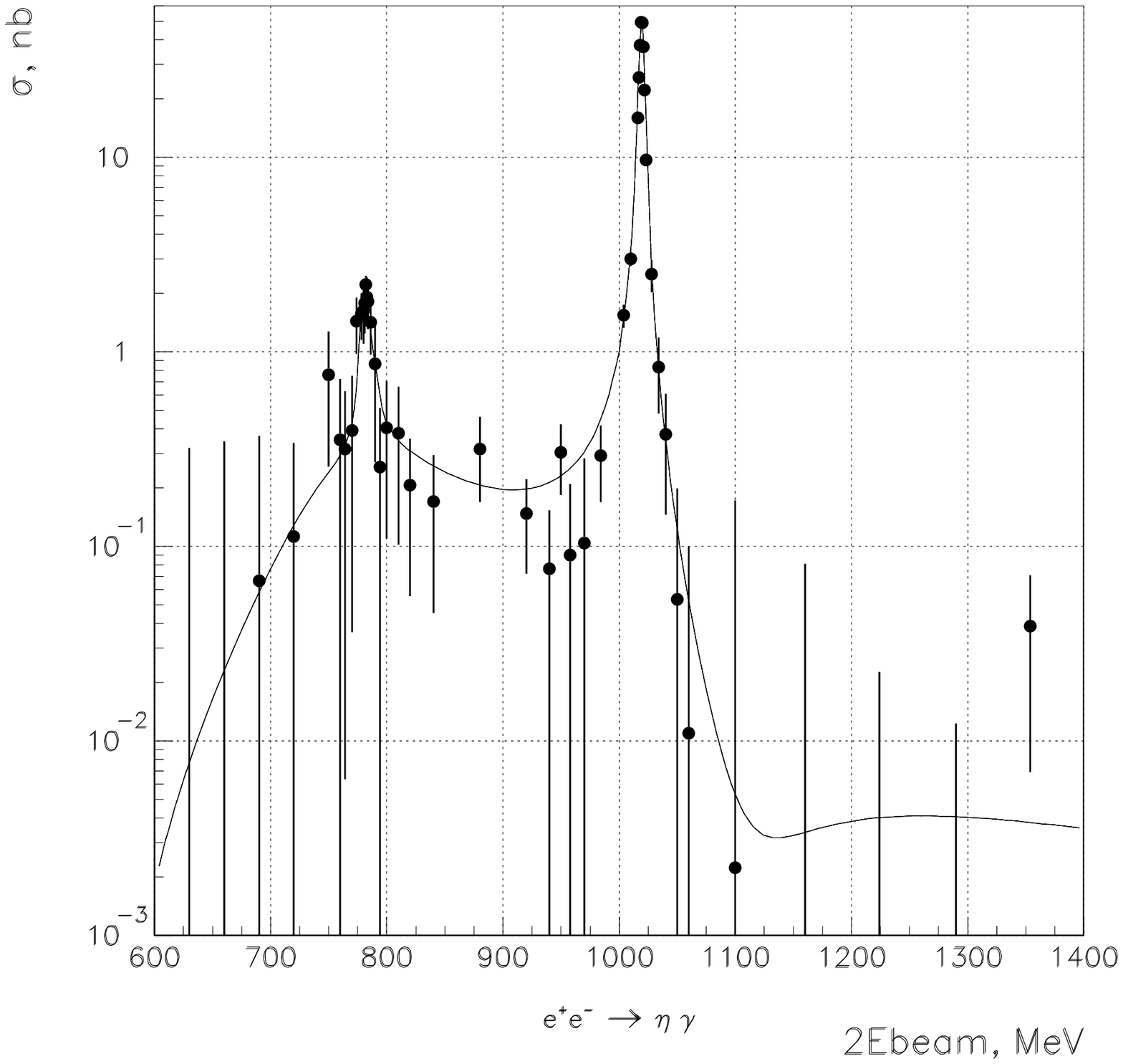}
\hfill
\includegraphics[width=0.47\textwidth]{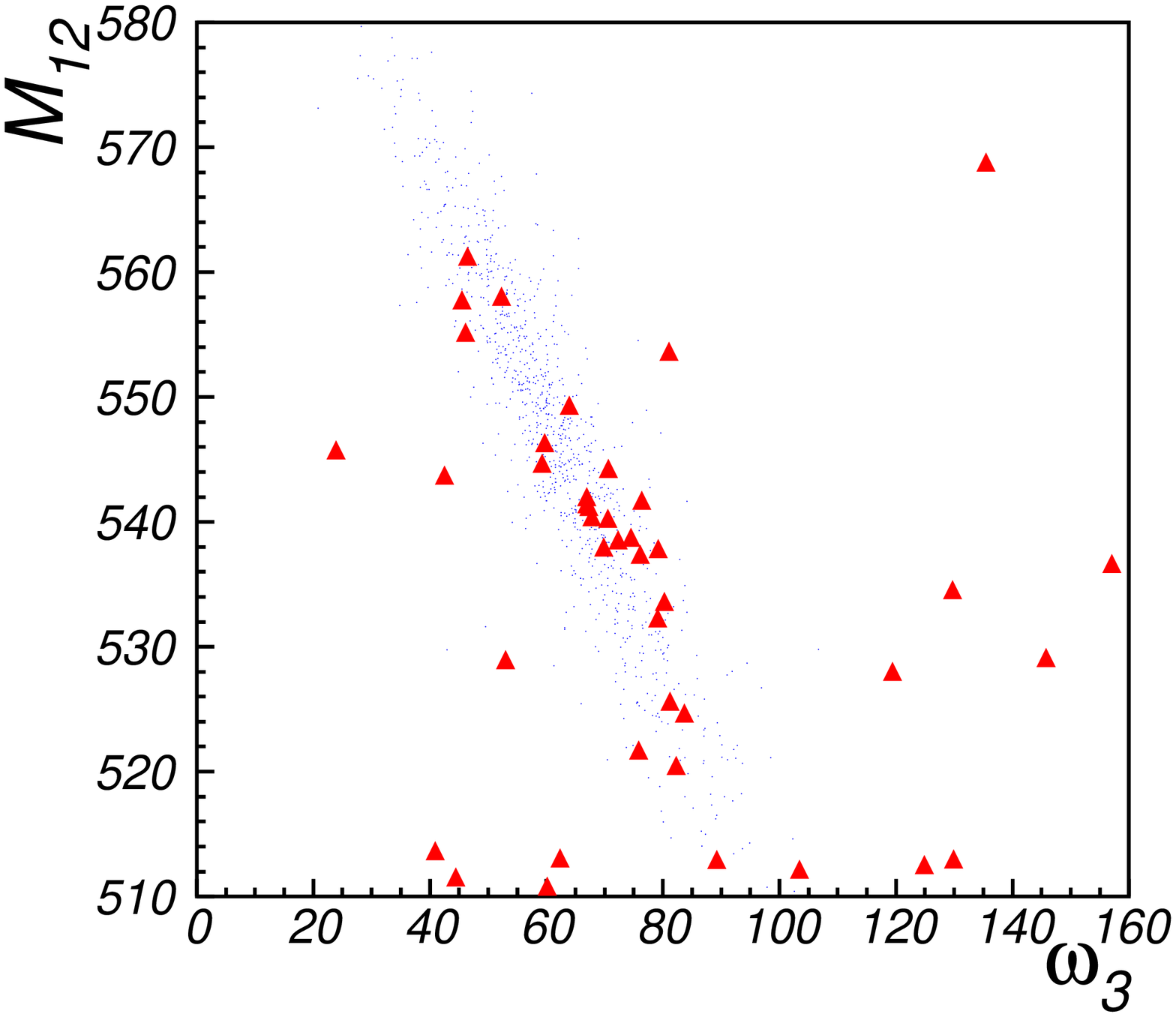}
\parbox[t]{0.47\textwidth}{
\caption{\it Cross section of the process $e^+e^- \to \eta\gamma
  \to 7 \gamma$}\label{fig:etag7g}
}
\hfill
\parbox[t]{0.47\textwidth}{
\caption{\it Invariant mass of two hard photons $M_{12}$ vs softest photon
energy $\omega_{3}$. Points present the simulation of $\phi \to
\eta' \gamma, \eta'\to\pi^+\pi^-\eta, \eta\to\gamma\gamma$, triangles ---
data after all the selections.}
\label{fig:finsum}
}
\end{figure}

The reaction $e^+e^- \to \eta\gamma$ when $\eta$ decays into
$3\pi^0$ has been studied in the energy range from 0.6 to 1.4 GeV with
the integrated luminosity about 21 $\mbox{pb}^{-1}$. The preliminary
results of measurement of the cross section of the process are
shown in fig.\ref{fig:etag7g}. The curve in this figure shows the fit of
the energy dependence of the cross section which takes into account
the interference of $\rho$, $\omega$ and $\phi$ mesons in the
intermediate state. The following branching ratios were obtained from
the fit: \\
{\boldmath
$Br(\phi\to\eta\gamma) = (1.24 \pm 0.02 \pm 0.08)\cdot 10^{-2}$, \\
$Br(\omega\to\eta\gamma) = (5.6^{+1.2}_{-1.1})\cdot 10^{-4}$, \\
$Br(\rho\to\eta\gamma) = (2.1^{+0.6}_{-0.5})\cdot 10^{-4}$
}

\section{Observation of $\phi\to\eta'\gamma$ decay}
   A search of this rare radiative decay was performed with the integrated 
luminosity of about 14 $\\mbox{pb}^{-1}$ at 14 energy points around
the $\phi$ meson  
when $\eta'$ decays into $\pi^+\pi^-\eta$.
The analysis of events has been performed using three different decay 
modes of $\eta$: 
a. $\eta\to\gamma\gamma$, b. $\eta\to
\pi^+\pi^-\gamma$ and  c. $\eta\to\pi^+\pi^-\pi^0$. 

   For the first case (a), 
there are two charged pions and three photons in the final state. The 
monochromatic recoil photon has a fixed energy of 60 MeV. The invariant mass
of two other (more hard) photons should equal $M_{\eta}$. 
Fig.\ref{fig:finsum} shows the distribution of the invariant mass of
two hard photons $M_{12}$ versus the softest photon energy $\omega_{3}$.
The main source 
of the background comes from the decay $\phi\to\eta\gamma$ when $\eta$
decay into $\pi^+\pi^-\pi^0$. In this case 
the final state has the same particles  but 
their kinematics is drastically different. The hardest photon is 
monochromatic with the energy 362 MeV and the invariant mass of two others
is $M_{\pi^0}$. The decay $\phi\to\eta\gamma$ is two orders of 
magnitude more probable.
The branching ratio of $Br(\phi\to\eta'\gamma)$ was calculated 
relative to $Br(\phi\to\eta\gamma)$. This ratio is not sensitive
to systematic uncertainties from 
luminosity, detector inefficiency, resolution and so on. Using the 
values of the all needed branching ratios from PDG\cite{PDG} the following 
result has been obtained: 
{\boldmath
$Br(\phi\to \eta'\gamma)=(0.82^{+0.21}_{-0.19}\pm 0.11)\times 10^{-4}$.
}    

   For the second(b) and third(d) decay modes of $\eta$ there are four charged 
particles and two or three photons in the final state. The softest photon
is monochromatic with the energy of 60 MeV. One of the  combinations of 
two particles with opposite charges has to form a missing mass to 
$M_{\pi^0}$ or zero. The kinematic constrained fit with additional 
angular cuts was applied to select events with the best $\chi^2$. The 
main source of the background comes from decays: $\phi\to 
K^{0}_S K^{0}_L$
when $K^{0}_S\to\pi^+\pi^-$ and $K_L\to\pi^+\pi^-\pi^0$. 
The number of these background events was subtracted according to 
the simulation results. The following result has been obtained :
{\boldmath
$Br(\phi\to \eta'\gamma)=(0.58\pm 0.18\pm 0.15)\times 10^{-4}$.}

\section{Direct observation of $K^{0}_{S}\to\pi e \nu$ decay}
    While the semileptonic decays of $K^{0}_{L}$ have been well measured, 
the information on the similar decays of $K^{0}_{S}$ is extremely scarce.
PDG evaluates the corresponding decay rate indirectly using the $K^{0}_{L}$
semileptonic decays and assuming the rule: $\Delta$S=$\Delta$Q.

\begin{figure}[ht]
\begin{center}
\includegraphics [width=0.47\textwidth]{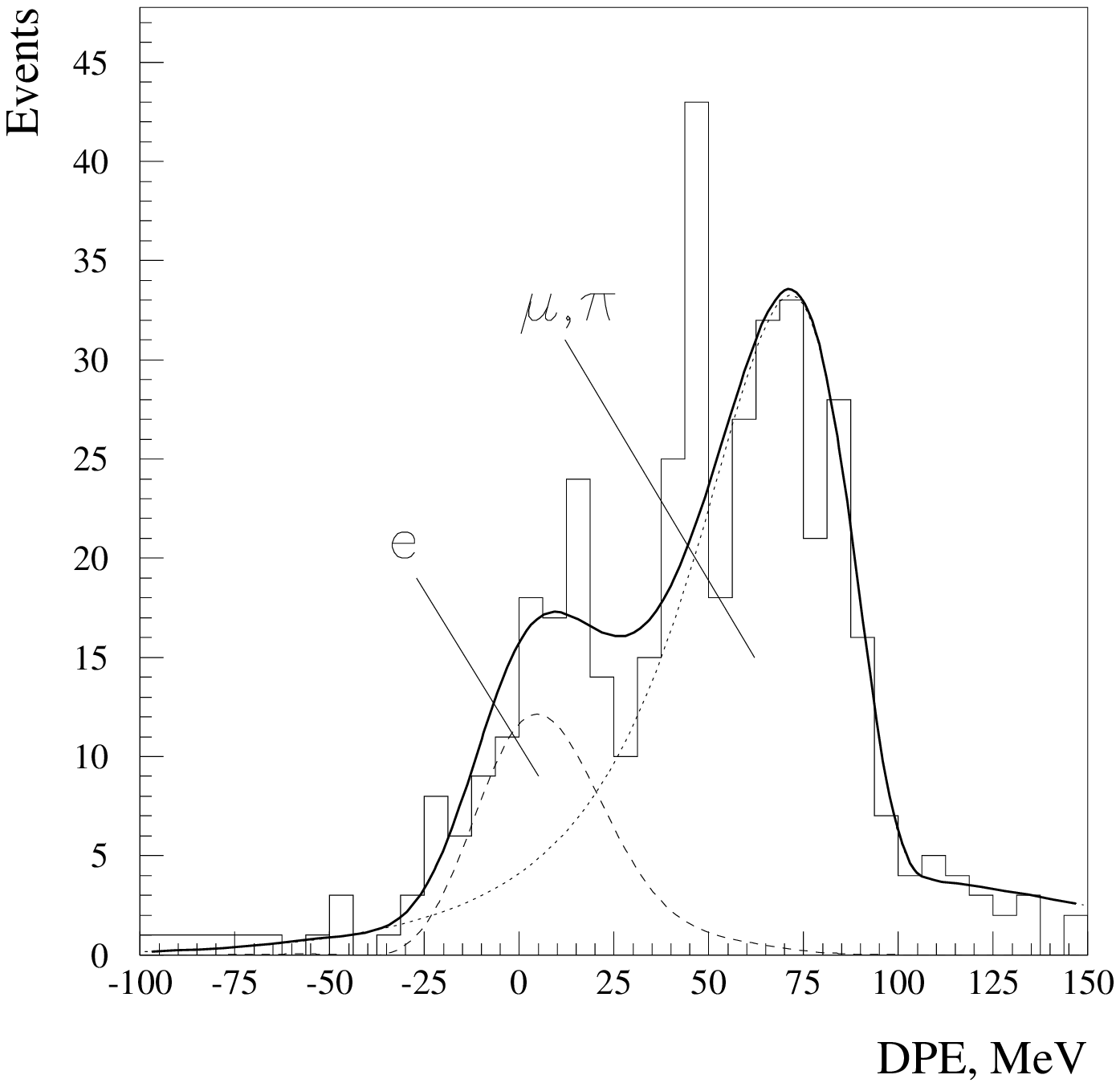}
\end{center}
  \caption{\it DPE distribution for charged particles in $K_S^0$ decays;
    dashed line --- electrons; dotted line --- muons and pions}
  \label{picrys6}
\end{figure}

    We present results of the direct measurement of the branching ratio
for  $K^{0}_{S}\to\pi e \nu$ decay using the unique opportunity 
to study events containing a pure $K^{0}_{L} K^{0}_{S}$ system in the final
state produced in the reaction: 
$e^+e^-\to\phi\to K^{0}_{L} K^{0}_{S}$. 
The data with the integrated luminosity of 14.8 $\mbox{pb}^{-1}$ were used for
this analysis.
Fig.\ref{picrys6} shows the distribution of the selected events
over the parameter $DPE=p-E_{loss}-E_{CsI}$, where $p$ is the particle
momentum measured in drift chamber, $E_{CsI}$ is the energy deposition
in CsI calorimeter and $E_{loss}$ is the an ionization energy loss in
the material in front of CsI calorimeter. An enhancement in this
distribution around zero corresponds to the electrons from the decay
$K^{0}_{S}\to \pi e \nu$. After
background subtraction the corresponding number of the events was found
to be: $ N = 75\pm 13 $. Using the $K^{0}_{S}\to\pi^+\pi^-$ decay
for normalization, the following branching ratio was obtained:
{\boldmath
$Br(K^{0}_{S}\to\pi e \nu)$ = $(7.19 \pm 1.35)\times 10^{-4}$.}
This result is consistent with the PDG value obtained by recalculation from 
$K^{0}_{L}$ semileptonic rates. More details on the analysis can be
found in\cite{ks-pen}.

\section{Study of the conversion decays}   
\providecommand{\etagg}{\ensuremath{\mbox{$\eta\to\gamma\gamma$}}}
\providecommand{\piogg}{\ensuremath{\mbox{$\pi^0\to\gamma\gamma$}}}
\providecommand{\etapz}{\ensuremath{\mbox{$\eta{\to}3\pi^0$}}}
\providecommand{\phietaee}{\ensuremath{\mbox{$\phi\to{\eta}e^+e^-$}}}
\providecommand{\etaeeg}{\ensuremath{\mbox{$\eta{\to}e^+e^-\gamma$}}}
\providecommand{\phipioee}{\ensuremath{\mbox{$\phi\to{\pi^0}e^+e^-$}}}
\providecommand{\phietag}{\ensuremath{\mbox{$\phi\to\eta\gamma$}}}
\providecommand{\etagppg}{\ensuremath{\mbox{$\phi\to\eta\gamma$},
\ \mbox{$\eta\to\pi^+\pi^-\gamma$}}}
\providecommand{\phippp}{\ensuremath{\mbox{$\phi\to\pi^+\pi^-\pi^0$}}}
   Conversion decays, when a virtual photon is converted into a lepton
pair, are closely related to corresponding radiative decays.  
The branching ratios for conversion decays \phietaee, \phipioee\   
as well as Dalitz decay \etaeeg\  were determined using a data sample 
with the integrated luminosity of 15.5 $\mbox{pb}^{-1}$. 

   The decay \phietaee\   was detected via the mode \etagg\  
and \etapz, 
the decay \phipioee\  --  via the \piogg\   
and the decay \etaeeg\  -- via the mode \phietag.
The process \etagppg\   was used to determine the number of $\phi$-mesons.

   Events were selected with two charged particles in DC and photons
in the calorimeter. These events were subject to the kinematic fit  
with energy-momentum conservation.
The conversion decays have a peculiar feature of their
kinematics: the angle between $e^+$ and $e^-$ is as a rule close to zero.

\begin{figure}[ht]
\begin{center}
\includegraphics[width=0.47\textwidth]{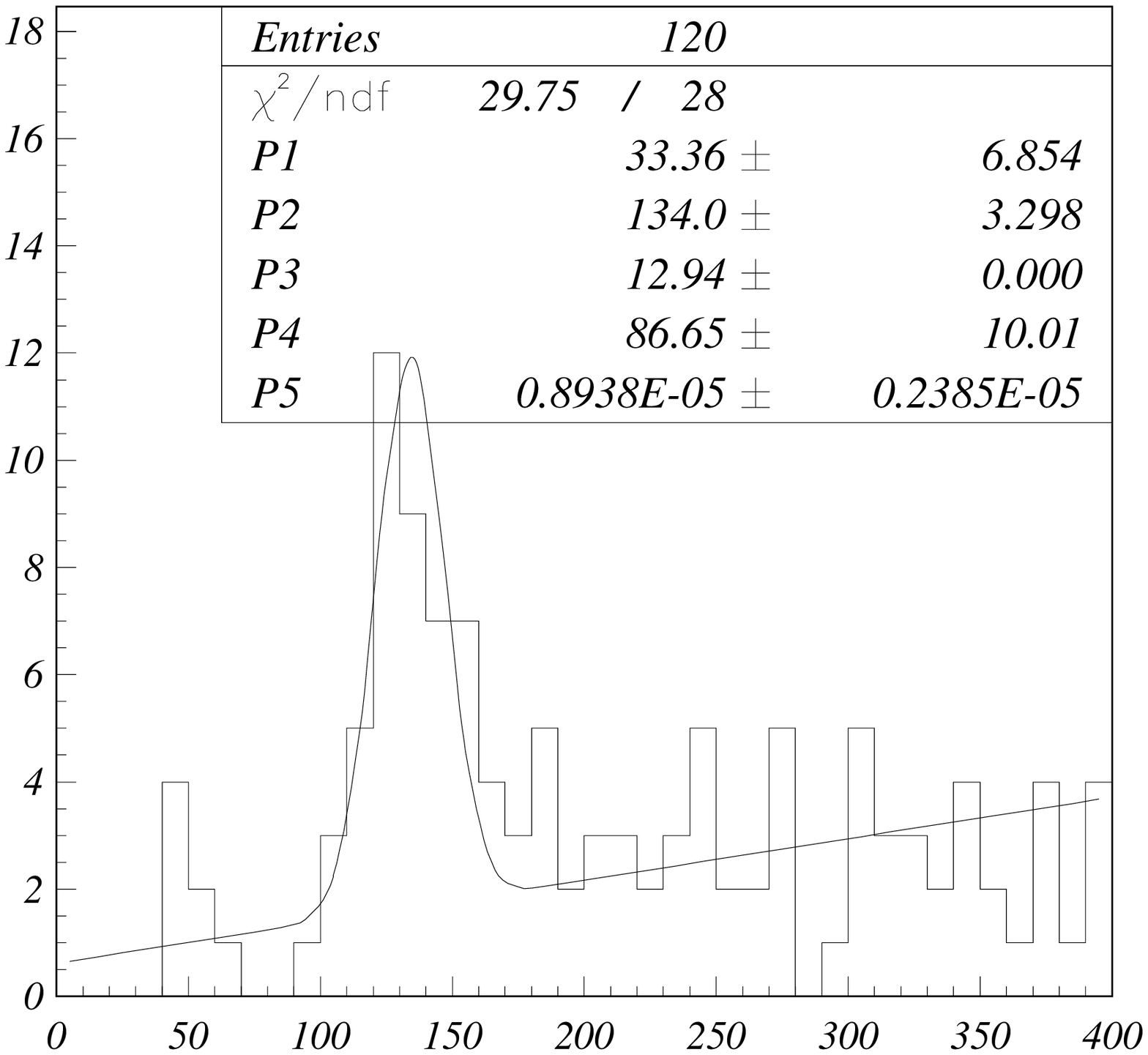}
\end{center}
\caption{\it Distribution over the invariant mass of pair of photons
  for the process $\phi\to \pi^0 e^+ e^-$} \label{fig:pi0ee}
\end{figure}

   The significant background for these events comes from
the $\gamma$-quantum conversion in the detector material. 
The detection efficiencies for these processes were determined by simulation. 
The decay \phipioee\  has background from \phippp\  via the same final state. 
This background was suppressed by using the information about   
energy deposition by electrons and pions in the calorimeter.
Fig.\ref{fig:pi0ee} shows the distribution over the invariant mass of
pair of photons for the events of the process $\phi\to\pi^0
e^+e^-$.
As a preliminary result the following branching ratios were obtained:\\
{\boldmath
$Br(\phi\to\eta e^+e^-)$ = $(1.01\pm0.14\pm0.15)\times 10^{-4}$} when 
$\eta\to\gamma\gamma$,\\
{\boldmath
$Br(\phi\to\eta e^+e^-)$ = $(1.20\pm0.22\pm0.18)\times 10^{-4}$} when 
$\eta\to\pi^0\pi^0\pi^0$,\\
{\boldmath
$Br(\phi\to\pi^{0} e^+e^-)$ = $(1.23\pm0.33\pm0.20)\times 10^{-5}$,
}
{\boldmath
$Br(\eta\to e^+e^-\gamma)$ = $(6.85\pm0.60\pm1.00)\times 10^{-3}$.}
  The obtained results are in agreement with the theoretical predictions
and have better statistical accuracy than 
previous measurements quoted by PDG.

\section{Reactions \eefourpi\ and $e^+e^-\to \pne$}

\begin{figure}[ht]
\includegraphics[width=0.47\textwidth]{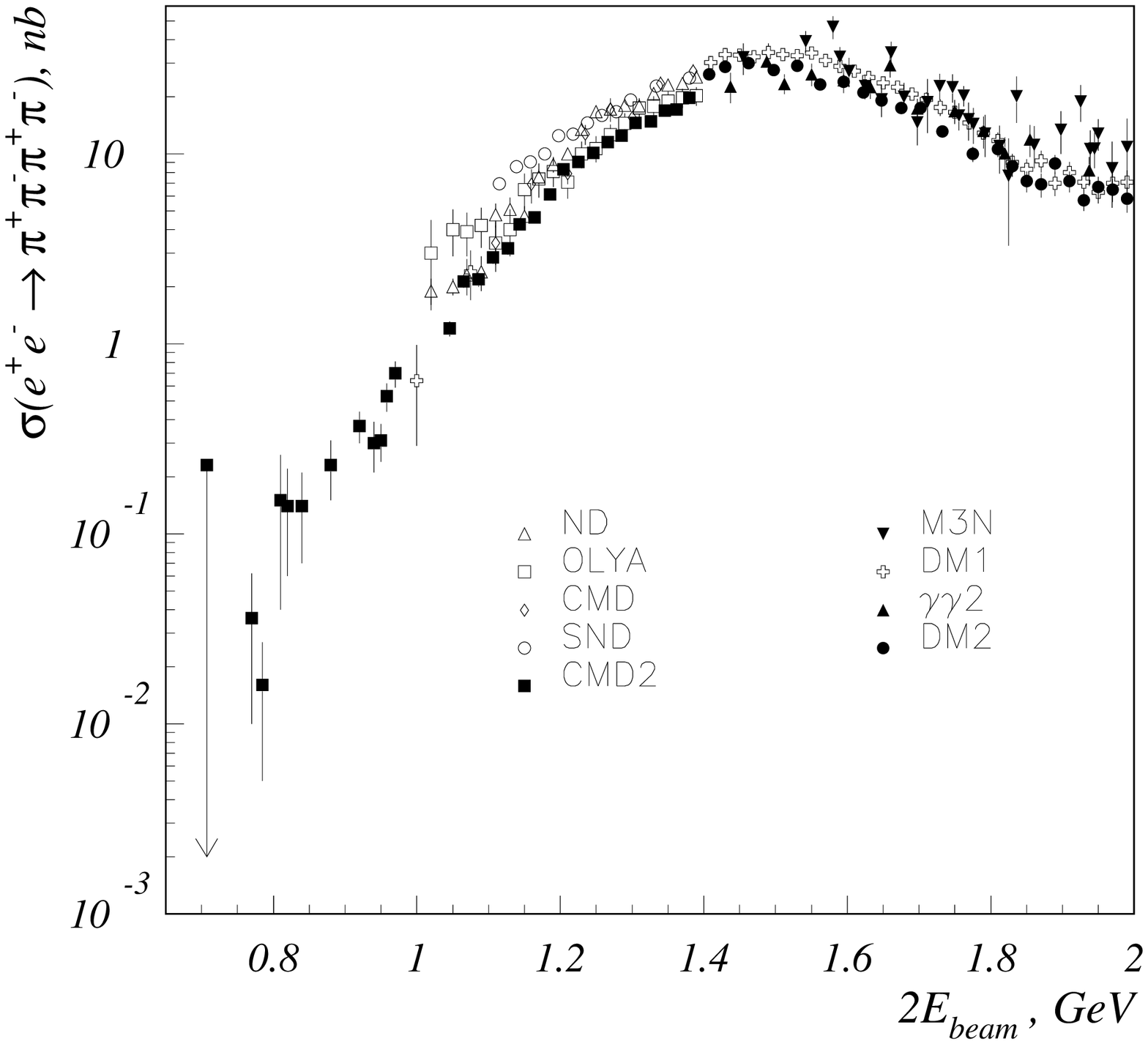}
\hfill
\includegraphics[width=0.47\textwidth]{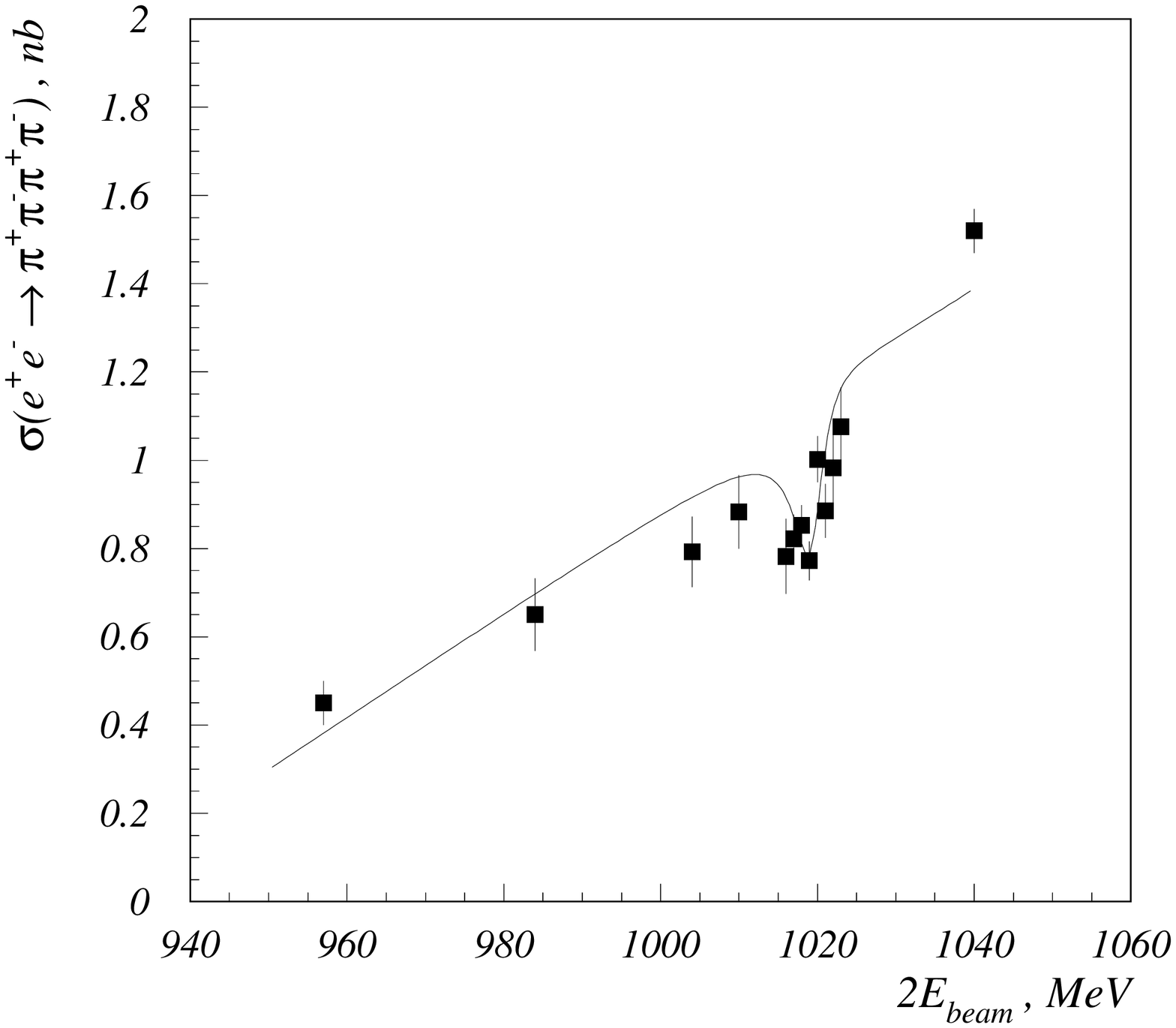}
\parbox[t]{0.47\textwidth}{\caption{\it Cross section of the reaction
    \eefourpi}\label{fig:xsec-4pi}}
\hfill
\parbox[t]{0.47\textwidth}{\caption{\it Cross section of the process
    \eefourpi\ in the $\phi$ meson region}\label{fig:phi-4pi}}
\end{figure}
The reaction of $e^+e^-$ annihilation into four pions (with two
possible channels \fourpi\ and \pne) was studied in the energy range
1.05--1.38 GeV. Simultaneous analysis of both modes  allowed to
establish that the final state \pne\ is dominated by a mixture of
$\omega\pi^0$ and $a_1(1260)\pi$ mechanisms whereas only the latter
contributes to the \fourpi\ final state\cite{4pi-high}.
The reaction \eefourpi\ was also studied in the energy range 0.6--0.97
GeV\cite{4pi-low}. The energy dependence of the cross section in this
range agrees with the assumption of the $a_1(1260)\pi$ intermediate
state.
Fig.\ref{fig:xsec-4pi} shows the energy behavior of the cross section
of the reaction \eefourpi\ in the energy range 0.6 to 2 GeV. Also
shown in this figure are the measurements of other groups.
For the first time \fourpi\ events were observed at the $\rho$ meson
energy. Under the assumption that all these events come from $\rho$
meson decay, the following value of the decay width was obtained:\\
{\boldmath
$\Gamma(\rho^0\to\fourpi) = (2.8\pm 1.4\pm 0.5)$ } keV \\
or the branching ratio: \\
{\boldmath
$Br(\rho^0 \to \fourpi) = (1.8 \pm 0.9 \pm 0.3)\cdot 10^{-5}$
}

Fig.\ref{fig:phi-4pi} shows the preliminary results of measurement of
the cross section of the process \eefourpi\ near $\phi$ meson. A
signal of the decay $\phi\to \fourpi$ is well seen in this
figure. The following branching ratio was obtained from the fit: \\
{\boldmath
$Br(\phi\to\fourpi) = (5.4 \pm 1.6 \pm 2.0)\cdot 10^{-6} $
}

\section{Reaction $e^+e^-\to\pi^+\pi^-\pi^+\pi^-\pi^0$}

\begin{figure}[ht]
\includegraphics[width=0.47\textwidth]{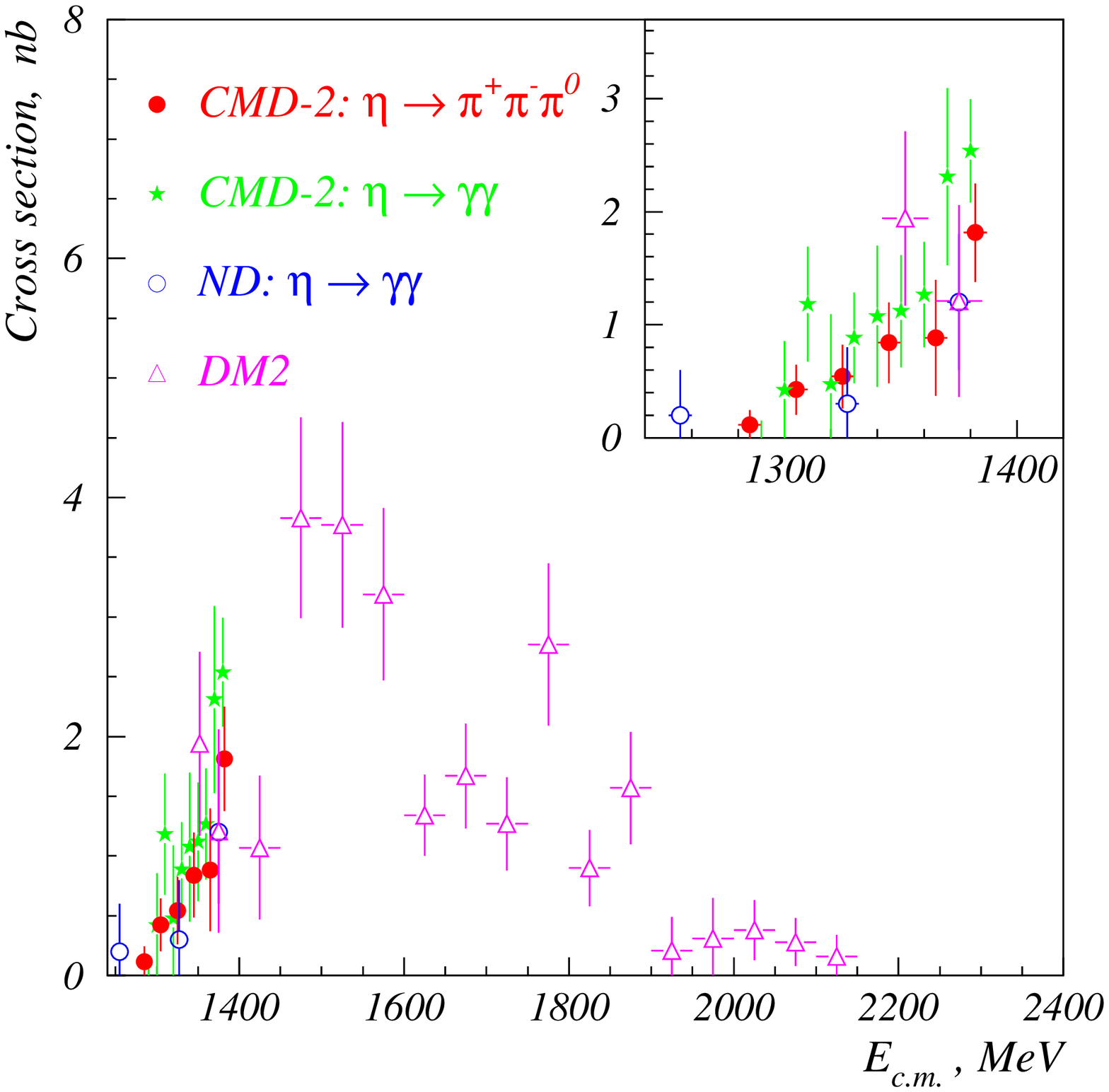}
\hfill
\includegraphics[width=0.47\textwidth]{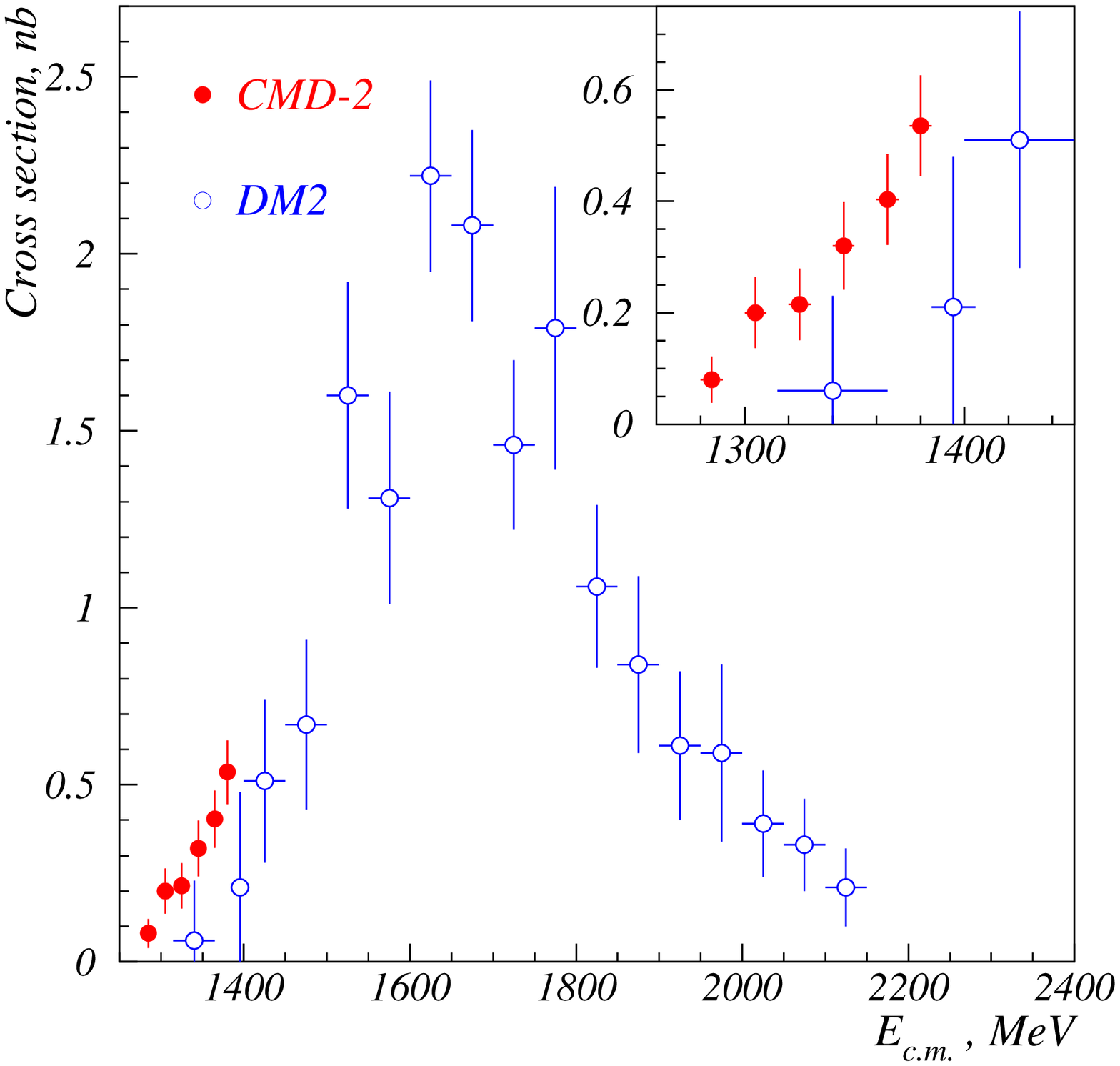}
\parbox[t]{0.47\textwidth}{\caption{\it Cross section of the reaction
    $e^+e^-\to\eta\pi^+\pi^-$}\label{fig:etapipi}}
\hfill
\parbox[t]{0.47\textwidth}{\caption{\it Cross section of the process
    $e^+e^-\to\omega\pi^+\pi^-$}\label{fig:ompipi}}
\end{figure}

The energy dependence of the cross section of the
process $e^+e^-\to\pi^+\pi^-\pi^+\pi^-\pi^0$ was
measured. The dominance by the contributions from the $\eta\pi^+\pi^-$
and $\omega\pi^+\pi^-$ states was shown. The reaction
$e^+e^-\to \eta\pi^+\pi^-$ was also studied when $\eta$ decays
into $\gamma\gamma$. The results of measurements are shown in
fig.\ref{fig:etapipi} and \ref{fig:ompipi}.

\section{Conclusion}

New interesting results were obtained with CMD-2 detector on VEPP-2M
collider. Among them are the high precision measurement of the cross
section of $e^+e^-$ annihilation into hadrons in the energy range from
the threshold of hadron production to 1.4 GeV, investigation of exclusive
hadron channels as well as decays of $\rho$, $\omega$ and $\phi$ mesons. 
Some of these results are outlined in the tab.\ref{tab:cmd2res}.
Analysis is in progress to produce final results with a low systematic
uncertainty to meet the original goals of CMD-2.

\begin{table}
\caption{\it Results of CMD-2 experiments}\label{tab:cmd2res}
\begin{center}
\begin{tabular}{|l|c|c|}
\hline
& CMD-2 Data & PDG'98 \\
\hline
$m_{\omega}$, MeV & $782.71 \pm 0.07 \pm 0.04$ & $781.94 \pm 0.12$ \\
$\Gamma_{\omega}$, MeV & $8.68 \pm 0.23 \pm 0.20$ & $8.41 \pm 0.09$ \\
$\Gamma_{\omega\to e^+e^-}$, keV &
$0.605 \pm 0.014 \pm 0.010$ & $0.60 \pm 0.02$ \\
$Br(\omega\to\pi^+\pi^-)$ &
$(1.31 \pm 0.23 \pm 0.02)\cdot 10^{-2}$ & $(2.21 \pm 0.30)\cdot 10^{-2}$ \\
$m_{\rho}$, MeV & $775.28 \pm 0.61 \pm 0.20$ & $776.0 \pm 0.9$ \\
$\Gamma_{\rho}$, MeV & $147.70 \pm 1.29 \pm 0.40$ & $150.5 \pm 2.7$ \\
$\Gamma_{\rho\to e^+e^-}$, keV &
$6.93 \pm 0.11 \pm 0.10$ & $6.77 \pm 0.32$ \\
$Br(\rho \to \pi^+\pi^-\pi^+\pi^-)$ &
$(1.8 \pm 0.9 \pm 0.3)\cdot 10^{-5}$ & $<2\cdot 10^{-4}$ \\
$Br(\rho \to \eta\gamma)$, $\eta\to 3\pi^0$ &
$(2.1^{+0.6}_{-0.5})\cdot 10^{-4}$ & $(2.4^{+0.8}_{-0.9})\cdot 10^{-4}$ \\
$Br(\omega \to \eta\gamma)$, $\eta\to 3\pi^0$ &
$(5.6^{+1.2}_{-1.1})\cdot 10^{-4}$ & $(6.5 \pm 1.0)\cdot 10^{-4}$ \\
\hline
$Br(K^{0}_{S}\to \pi e \nu)$ & $(7.2 \pm 1.4)\cdot 10^{-4}$ &
$(6.70 \pm 0.07) \cdot 10^{-4}$ \\
$Br(K^+ \to \pi^+\pi^0)$ & $(21.69 \pm 0.48 \pm 1.03)\cdot 10^{-2}$ & 
$(21.16 \pm 0.14)\cdot 10^{-2}$ \\
$Br(K^+ \to \pi^0 e^+ \nu)$ & $(4.89 \pm 0.17 \pm 0.17)\cdot 10^{-2}$ &
$(4.82 \pm 0.06)\cdot 10^{-2}$ \\
$Br(\eta\to e^+e^-\gamma)$ & $(6.85 \pm 0.60 \pm 0.90)\cdot 10^{-3}$ &
$(4.9 \pm 1.1)\cdot 10^{-3}$ \\
$Br(\eta \to \pi^+\pi^-e^+e^-)$ &  $(3.5 \pm 2.0)\cdot10^{-4}$  & 
$(13^{+12}_{-8})\cdot10^{-4}$ \\
$Br(\eta \to \pi^+\pi^-)$ &  $<3.3\cdot10^{-4}$  & $<9\cdot10^{-4}$ \\
$Br(\eta \to \pi^0\pi^0)$ &  $<5\cdot10^{-4}$  & no data \\
\hline
$m_{\phi}$, MeV & $1019.470 \pm 0.013 \pm 0.018$ & $1019.413 \pm
0.008$ \\
$\Gamma_{\phi}$, MeV & $4.51 \pm 0.04 \pm 0.02$ & $4.43 \pm 0.05$ \\
$Br(\phi\to \eta\gamma)$, $\eta\to 3\pi^0$ &
$(1.24 \pm 0.02 \pm 0.08)\cdot 10^{-2}$ & $(1.26 \pm 0.06)\cdot 10^{-2}$ \\
$Br(\phi\to \eta\gamma)$, $\eta\to\pi^+\pi^-\pi^0$ &
$(1.18 \pm 0.03 \pm 0.06)\cdot 10^{-2}$ & $(1.26 \pm 0.06)\cdot 10^{-2}$ \\
$Br(\phi\to e^+e^-)$ & $(2.87 \pm 0.09)\cdot 10^{-4}$ &
$(2.99 \pm 0.08)\cdot 10^{-4}$ \\
$\delta_{\phi - \omega}$ & $(162 \pm 17)^{\circ}$ & \\
$Br(\phi\to\eta^{\prime}\gamma)$, $\eta^{\prime}\to
\pi^+\pi^-\gamma\gamma$ &
$(8.2^{+2.1}_{-1.9} \pm 1.1)\cdot 10^{-5}$ & $(12^{+7}_{-5})\cdot
10^{-5}$ \\
$Br(\phi\to\eta^{\prime}\gamma)$, $\eta^{\prime}\to\fourpi\pi^0$ &
$(5.8 \pm 1.8 \pm 1.5)\cdot 10^{-5}$ & $(12^{+7}_{-5})\cdot 10^{-5}$ \\
\multicolumn{1}{|r|}{or $\eta^{\prime}\to\fourpi\gamma$} & & \\
$Br(\phi\to \eta e^+e^-)$, $\eta\to\gamma\gamma$ &
$(1.01 \pm 0.14 \pm 0.15)\cdot 10^{-4}$ & $(1.3^{+0.8}_{-0.6})\cdot
10^{-4}$ \\
$Br(\phi\to \eta e^+e^-)$, $\eta\to\pi^+\pi^-\pi^0$ &
$(1.00 \pm 0.18)\cdot 10^{-4}$ & $(1.3^{+0.8}_{-0.6})\cdot
10^{-4}$ \\
$Br(\phi\to \eta e^+e^-)$, $\eta\to 3\pi^0$ &
$(1.20 \pm 0.22 \pm 0.18)\cdot 10^{-4}$ & $(1.3^{+0.8}_{-0.6})\cdot
10^{-4}$ \\
$Br(\phi\to \pi^0 e^+e^-)$ & $(1.40 \pm 0.33 \pm 0.21)\cdot 10^{-5}$ &
$< 1.2 \cdot 10^{-4}$ \\
$Br(\phi \to \mu^+\mu^-\gamma)$ & $(1.43\pm 0.45\pm 0.14)\cdot 10^{-5}$ &  
$(2.3\pm 1.0)\cdot10^{-5}$ \\
$Br(\phi \to \rho\gamma)$ & $<1.2\cdot10^{-5}$  &  $<7\cdot10^{-4}$\\
$Br(\phi \to \rho\gamma\gamma)$ & $<5\cdot 10^{-4}$  & no data \\
$Br(\phi \to \eta\pi^+\pi^-)$ & $<3\cdot 10^{-4}$  & no data \\
$Br(\phi \to \pi^+\pi^-\pi^+\pi^-)$ & $(5.4 \pm 1.6 \pm 2.0)\cdot 
10^{-6}$ & $<8.7\cdot 10^{-4}$ \\
\hline
\end{tabular}
\end{center}
\end{table}

   This work is supported in part by the grants: RFBR-98-02-17851,  
RFBR-99-02-17053, RFBR-99-02-17119, INTAS 96-0624.

\end{document}